\documentclass[aps,prapplied,twocolumn,groupedaddress]{revtex4-2}

\usepackage{graphicx}
\usepackage{hyperref}
\usepackage{color}
\usepackage{lineno}
\usepackage{multirow}
\hypersetup{hypertex=true, colorlinks=true, linkcolor=blue, anchorcolor=blue, citecolor=blue}

\begin{document}

\title{Realizing an entanglement-based multi-user quantum network with integrated photonics}

\author{Wenjun Wen, Zhiyu Chen, Liangliang Lu, Wenhan Yan, Wenyi Xue, Peiyu Zhang, Yanqing Lu, Shining Zhu, Xiao-song Ma$^{\ast}$}

\affiliation{National Laboratory of Solid-state Microstructures, School of Physics, College of Engineering and Applied Sciences, Collaborative Innovation Center of Advanced Microstructures, Nanjing University, Nanjing 210093, China\\
$^{\ast}$e-mails: Xiaosong.Ma@nju.edu.cn}

\date{\today}

\begin{abstract}
	Quantum networks facilitate the secure transmission of information between different users. Establishing communication links among multiple users in a scalable and efficient way is important for realizing a large-scale quantum network. Here we develop an energy-time entanglement-based dense wavelength division multiplexed network based on an integrated silicon nitride microring resonator, which offers a wide frequency span (covering at least the entire C-band) and narrow bandwidth modes ($ \sim 650{\rm MHz} $). Six pairs of photons are selected to form a fully and simultaneously connected four-user quantum network. The observed quantum interference visibilities are well above the classical limits among all users. Each pair of users perform the BBM92 protocol for quantum key distribution. Our results pave the way for realizing large-scale quantum networks with integrated photonic architecture.
\end{abstract}

\maketitle

\section{INTRODUCTION}

Quantum entanglement, a counterintuitive feature of quantum physics, is the essential resource of quantum communication \cite{Gisin2002}, quantum computation \cite{Ladd2010}, and quantum metrology \cite{Giovannetti2011}. As quantum information carriers, photons have various degrees of freedom (DOFs) \cite{Thew2004, Dada2011, Richart2012, Wang2015, Schaeff2015, Kues2017, Imany2018, Wang2018, Reimer2019, Lu2020}. In particular, the frequency DOF of a single photon is attractive due to its compatibility with standard optical ﬁber infrastructure, and ability to perform routing based on optical frequencies \cite{Joshi2018}. Multiplexing the frequency DOF of photons is not only an efficient approach towards high data rates in classical communications \cite{Marin2017} but also a promising way to provide connectivity for quantum networks (QNs) \cite{Wengerowsky2018, Joshi2020, Liu2020, Liu2022, Erik2022}.

The development of QNs is essential for secure communication and information transfer among a larger number of users \cite{Wehner2018}. To date, QNs based on trusted relays or routing have been widely deployed \cite{Xu2009, Peev2009, Sasaki2011, Stucki2011, Chen2021}, and entanglement-swapping-based quantum direct communication network have also been demonstrated \cite{Li2019, Qi2021}. Fully and simultaneously connected QN schemes have been proposed and realized based on entanglement distribution without relying on trusted nodes \cite{Wengerowsky2018, Joshi2020, Liu2020, Liu2022}. In these pioneering studies, each pair of users share a bipartite entanglement through carving the broadband signal/idler spectrum into a series of slices and multiplexing them. However, as the number of users increases, the need for available wavelength channels increases, which is limited by the entangled photon spectrum and thus prevents the technology from extending to a larger scale. For a scalable QN, it is necessary to include quantum memories, relying on narrow-band optical transitions. Therefore, it is also highly desirable to realize a narrow-bandwidth entangled photon pair source. Moreover, one would like to have a stable, alignment-free, high-volume production solution for such sources in practice. By leveraging advances in complementary metal-oxide-semiconductor (CMOS) fabrication techniques for classical communication, integrated quantum photonics is a promising approach. Recent progress has focused on the generation of energy-time entangled optical frequency combs within a resonator \cite{Reimer2014, Mazeas2016, Jaramillo2017, Samara2019}. The discrete frequency modes generated from the resonators can easily extend the dimensionality of the system without significant overhead \cite{Kues2017, Imany2018, Reimer2016, Kues2019}.

\begin{figure*}[htbp]
	\centering
	\includegraphics[width=0.9\textwidth]{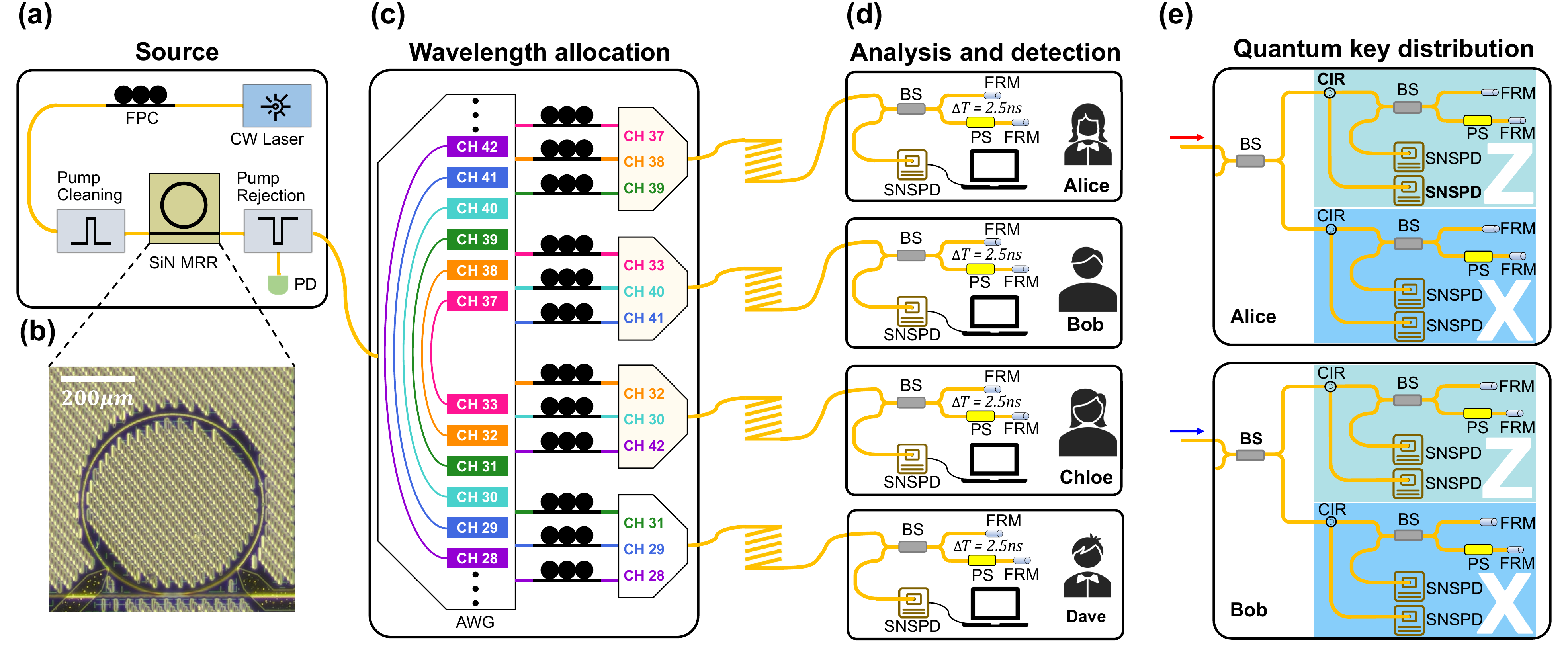}
	\caption{\label{Fig1}(a) Photon-pair source. A continuous wave laser at $ 1549.32 ~{\rm nm} $ is filtered, polarized and coupled into the Si$_3$N$_4$ microring resonator (MRR) to generate frequency-correlated photon pairs. At the output, the pump laser is rejected by a set of filters and measured by a power detector to monitor the source alignment. (b) Optical microscopy image of the Si$_3$N$_4$ MRR. (c) Wavelength allocation. Six bipartite states are selected to create a fully connected network between four users. (d) Each user is equipped with a Franson interferometer ($ \Delta T = 2.5 ~{\rm ns} $, time difference between two arms) and a superconducting single-photon detector to characterize the state. (e) The measurement setup for quantum key distribution using BBM92 protocol. For obtaining the secure keys, Alice/Bob split her/his photons with a 50:50 beam splitter and use two Franson interferometers with the phase settings corresponding to mutually unbiased bases \textit{Z} and \textit{X}, respectively. Abbreviations of components: CW, continuous wave; FPC, fiber polarization controller; MRR, microring resonator; PD, power detector; AWG, arrayed waveguide grating; BS, beam splitter; FRM, Faraday rotation mirror; PS, phase shifter; CIR, circulator; SNSPD, superconducting nanowire single-photon detector.}
\end{figure*}

Here we realize a four-user fully connected QN by employing a Si$_3$N$_4$ microring resonator (MRR), which generates comb-like bipartite energy-time entangled states through the nonlinear process of spontaneous four-wave mixing (SFWM) \cite{Li2005, Ramelow2015, Samara2019, Samara2021}. There are three advantages of our scheme. First, advances in the fabrication techniques allow the dispersion to be flexibly engineered to be anomalous over a broad bandwidth, benefiting the generation of a broadband frequency comb covering S, C, and L bands with different frequency spacings matching with standard telecom filters \cite{Marin2017, Reimer2016}. Second, field enhancement offered by the MRR offers high-brightness photon pairs, reduces the pump power requirement, and directly provides discrete narrow-linewidth spectral modes rather than filtering the photons after their generation \cite{Kues2019, Xie2015}, which reduces the photon count rate drastically. Such narrow-linewidth is compatible with quantum memories, which are necessary components for realizing future scalable QNs \cite{Simon2010}. Meanwhile, the relatively high bandgap makes the silicon nitride free of suffering two-photon absorption as compared with silicon MRR. Third, the source is chip-based, which offers stable and alignment-free operations with mature packaging technique. It is also CMOS compatible and hence is cost effective \cite{Mazeas2016, Takesue2007, Fortsch2013, Zhang2016, Lu2016, Pasquazi2018}. From a technological perspective, the time-bin qubit is the leading contender for fiber networks because this encoding does not suffer from random polarization rotation in fibers. By harnessing these advantages, we experimentally establish a four-user QN by employing six pairs of high-quality energy-time entangled photon pairs and distributing them between all users. Each user analyzes and decodes the time-bin qubits with a Franson interferometer.

\section{EXPERIMENTAL SETUP AND SOURCE CHARACTERIZATION}
As shown in FIG.\ref{Fig1}, our setup includes four parts: photon-pair source, wavelength allocation, entanglement analysis and quantum key distribution. In the source part (FIG.\ref{Fig1}(a)), a single-mode fiber-coupled continuous wave (CW) laser (Santec TSL-770) at $ 1549.32 ~{\rm nm} $ pumps the MRR source. We use a fiber polarization controller to adjust the pump polarization and pump cleaning filters to remove unwanted spectral noise before injecting the pump light into the MRR. The MRR has a diameter of $ 460 ~{\rm \mu m} $, as shown in FIG.\ref{Fig1}(b). The MRR is evanescently coupled to a single-bus waveguide via a $ 300 ~{\rm nm} $ gap point coupler. Both the ring and waveguide have a cross-section of $ 1600 ~{\rm nm} $ wide and $ 800 ~{\rm nm} $ high. Light is coupled into and out from the chip by lens tapered fibers which is mounted on high-precision positioning mounts. The total insertion loss (for nonresonant frequency) of the chip is about $ 5 ~{\rm dB} $, including both input-output coupling loss and the propagation loss in the bus waveguide.

Frequency-correlated photons are generated from the integrated nonlinear MRR. The broadband phase matching SFWM enabled by anomalous dispersion of our device, combined with the cavity resonances, leads to the generation of a correlated signal and idler photon pairs that are symmetrically distributed to the pump laser. The spectra of the entangled photon pairs cover multiple resonances, and yield a frequency comb structure with a spacing given by the free spectral range (FSR) of the MRR. The photons are coupled out from the chip and are filtered by cascade filters. Then they are allocated to users to form a fully and simultaneously connected network by an array waveguide grating (AWG) and multiple dense wavelength division multiplexing (DWDM) filters with $ 100 ~{\rm GHz} $ spacing (FIG.\ref{Fig1}(c)). We select six pairs of entangled photons with wavelengths aligned with the International Telecommunication Union’s (ITU) grids. These six pairs of photons are then distributed to four users: Alice (A), Bob (B), Chloe (C), and Dave (D). Each user receives three channels/wavelengths and shares an entanglement state with every other user in the network. The analysis and detection module held by every user consists of a Franson interferometer \cite{Franson1989} with a time difference between two arms of $ \Delta T = 2.5 ~{\rm ns} $, and a superconducting nanowire single-photon detector (SNSPD, Quantum Opus). All interferometers are composed of fiber optic components and are buried in quartz sand and wrapped in thick insulation to mitigate the phase shifts caused by room temperature changes. As shown in FIG.\ref{Fig1}(d), the generated photons can either take a short ($ S $) or a long ($ L $) path in the interferometer. We select the case when both photons take the same path, which gives the bipartite entangled state $ \left | \psi  \right \rangle = \frac{1}{\sqrt{2}} \left ( \left | SS  \right \rangle + e^{ \phi_{s} + \phi_{i} } \left | LL  \right \rangle  \right )  $ shared between each user \cite{Li2017, Li2019}, where $ \phi_{s} $ ($ \phi_{i} $) is the phase of the signal (idler) set by the phase shifter.

\begin{figure}[htbp]
	\includegraphics[width=0.45\textwidth]{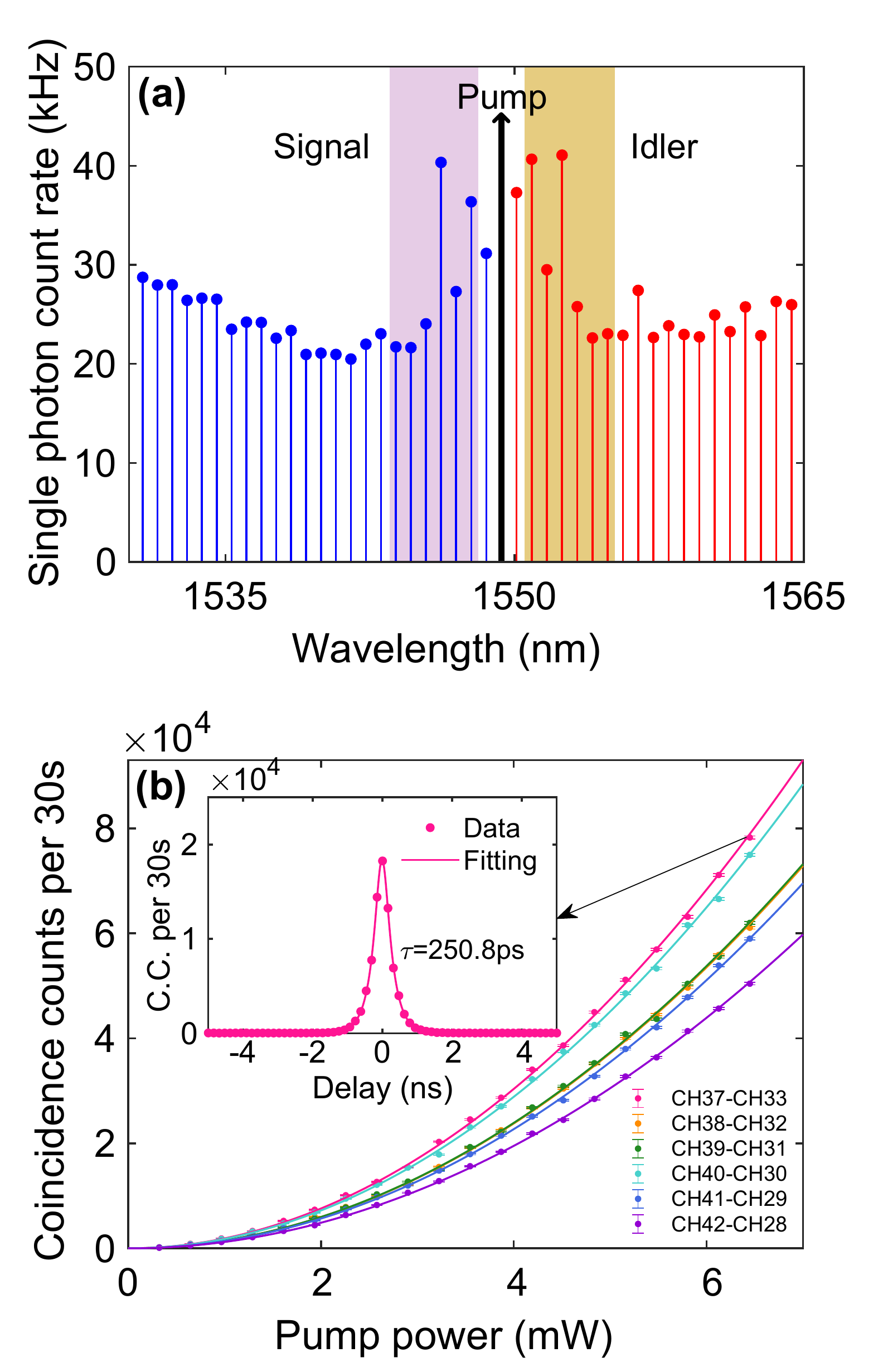}
	\caption{\label{Fig2}(a) Experimental single-photon spectrum as a function of the wavelength with the on-chip pump power of $ 5 ~{\rm mW} $. The channels used in the network are indicated in shades. (b) Coincidence counts as a function of pump power for six photon pairs of different frequencies with a coincidence window of $ 2.5 ~{\rm ns} $. Solid lines are quadratic fits. Error bars are calculated by Poissonian distribution and are smaller than the symbol size. The inset is a typical histogram of coincidence measurements with the time-bin size of $ 156.25 ~{\rm ps} $, which shows the coherence time of the single photons $ \tau_{c} $ of $ 250.8 ~{\rm ps} $.}
\end{figure}

The MRR under test has an average \textit{Q} factor of $ 3.1 \times 10^{5} $ and an average FSR of $  97.8 ~{\rm GHz} $. The full width at half maximum (FWHM) of the resonance mode near the pump wavelength is $ 5.2 ~{\rm pm} $ ($ 649 ~{\rm MHz} $). In FIG.\ref{Fig2}(a), we show the spectrum of the single photons emitted from the MRR and then measured with a reconfigurable optical processor (Finisar waveshaper 16000S-CB, not shown in the setup) and a SNSPD. Owing to the low anomalous dispersion of the MRR, we obtain the single-photon spectrum covering the entire C-band (only limited by our measurement spectrometer, not the comb itself). FIG.\ref{Fig2}(b) illustrates the dependence of coincidences on pump power for all six photon pairs with a coincidence window of $ 2.5 ~{\rm ns} $ measured directly from the output of the AWG. The generated rate of photon pairs shows a quadratic behavior with increasing pump power, which confirms the correlated photons are generated from the SFWM process. An instance of a histogram of the channel pairs (ITU CH37 and CH33) at the input pump power of $ 6.45 ~{\rm mW} $ is shown in the inset measured by a time-tagger unit (UQDevices Logic-16) with the temporal resolution of $ 156.25 ~{\rm ps} $. The peak represents the time correlation of the signal-idler pairs and a high coincidence-to-noise ratio (CAR, $ \sim 88 $). Following the fitting method used in \cite{Zhou2014, Reimer2014}, the results show that the coherence time of the single photons is $ \tau_{c} = 250.8 ~{\rm ps} $ and, hence, the bandwidth is $ \Delta \nu = 1/(2\pi\tau_{c}) = 634.7 ~{\rm MHz} $ (consistent with the resonance linewidth of the MRR), and the time jitter of the detection system (including SNSPD and time-tagger unit) is $ \sigma = 138.3 ~{\rm ps} $. The pair generation rate (PGR) can be calculated by $ PGR=\frac{S_sS_i}{R_c} $, where $ S_{s} $ ($ S_{i} $) is the single counts of the signal (idler) due to SFWM, and $ R_{c} $ is the coincidence count. After dividing by their respective bandwidth, the PGRs show a minimum of $ 11.2 ~{\rm s^{-1}mW^{-2}MHz^{-1}} $, a maximum of $ 15.3 ~{\rm s^{-1}mW^{-2}MHz^{-1}} $, and an average of $ 12.9 ~{\rm s^{-1}mW^{-2}MHz^{-1}} $. The details of the characterization are shown in Appendix A, and the results show that our source has the advantages of large spectral range, narrow linewidth of comb lines, high CAR, and high PGR, showing potential for applications in multiuser QNs and interfacing solid-state quantum memories at telecommunication wavelengths \cite{Saglamyurek2015}.

\section{RESULTS}
Each user receives three frequency modes, corresponding to three communication channels with the other three users in the four-user network. One way to distinguish the channel is to implement frequency-resolved detection at each user's detection module, which will drastically increase the complexity of the setup for each user. Instead, we employ the method used in ref \cite{Wengerowsky2018} by harnessing the temporal DOF. We add different optical fiber delays for each wavelength channel between wavelength demultiplexing and wavelength remultiplexing, i.e., directly behind the first AWG shown in FIG.\ref{Fig1}(c). By doing so, we can identify different combinations of users by the relative arrival times of respective photons.

\begin{figure}[htbp]
	\includegraphics[width=0.4\textwidth]{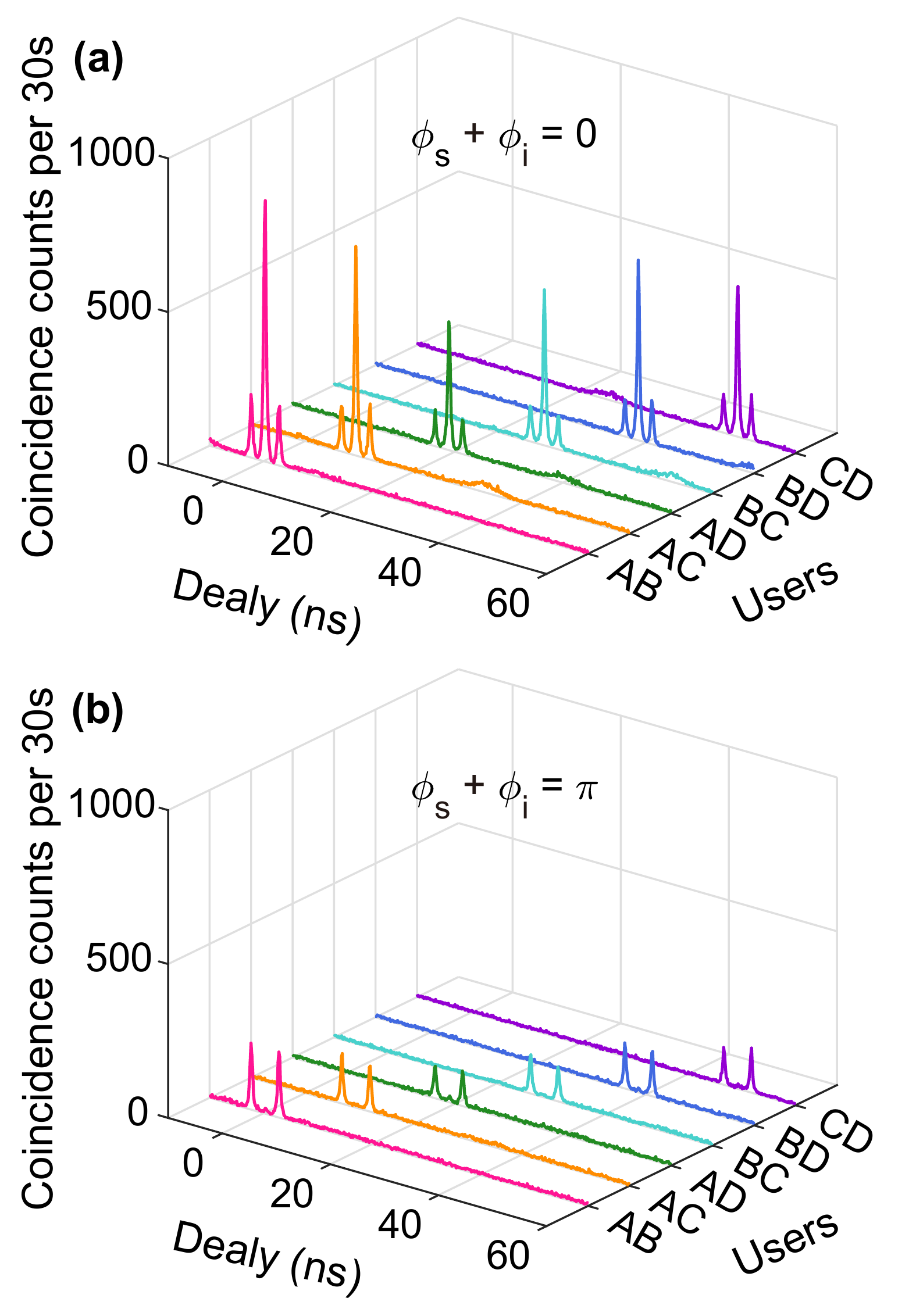}
	\caption{\label{Fig3}Coincidences histogram for each two-party link measured at the output of the interferometers for two different phase settings. Each pair of users identify photon pairs by different delays.}
\end{figure}

\begin{figure*}[htbp]
	\centering
	\includegraphics[width=0.9\textwidth]{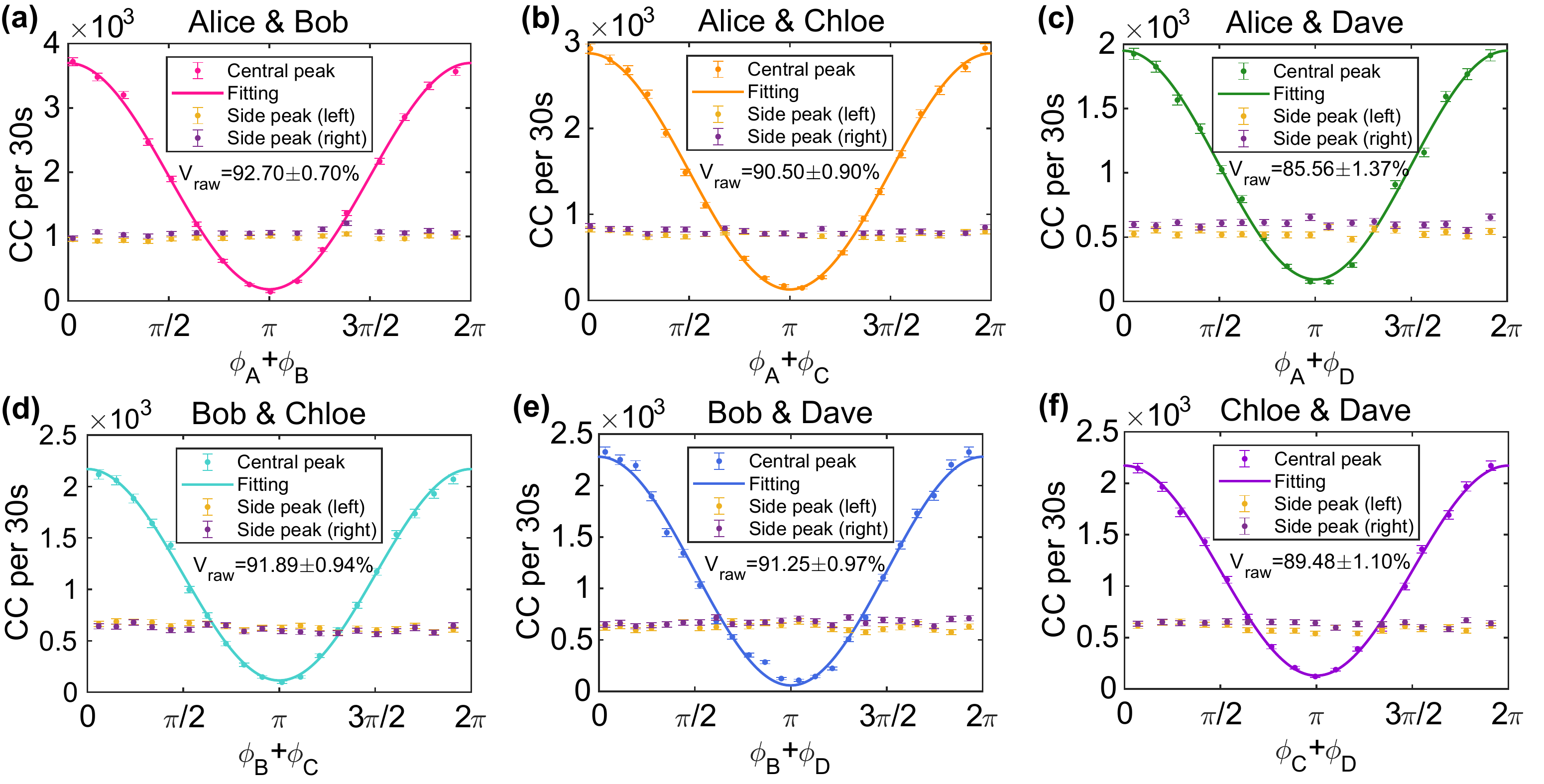}
	\caption{\label{Fig4}(a)-(f) Coincidence counts (CC) of the central and side peaks in $ 30 ~{\rm s} $ as a function of the phase $ \phi_{s} + \phi_{i} $ for six combinations among four users. All coincidence windows are $ 2.5 ~{\rm ns} $. The solid curves are fits of the experimental data of the central peak. Experimentally we scan the phase of the signal while keeping the idler’s phase without active tuning.}
\end{figure*}

\begin{figure*}[htbp]
	\centering
	\includegraphics[width=0.9\textwidth]{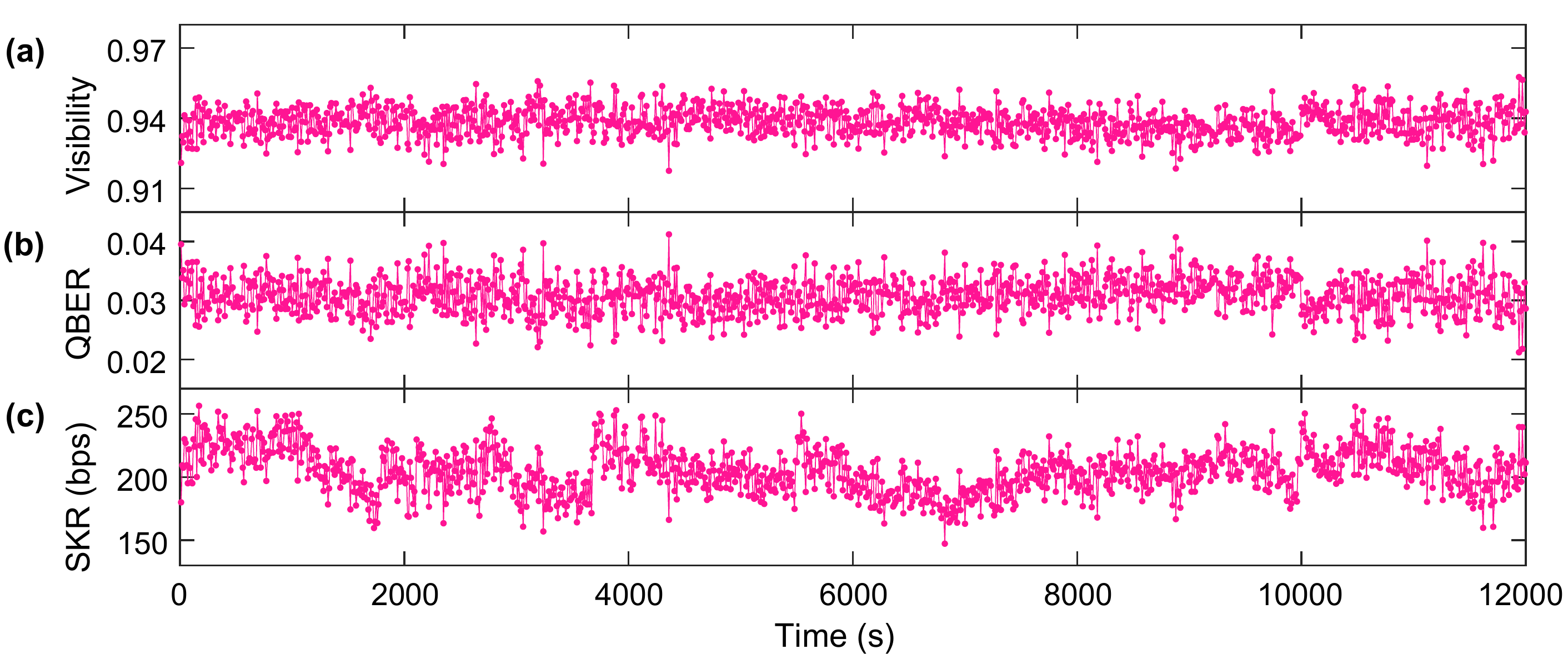}
	\caption{\label{Fig5}(a) Visibility, (b) quantum bit error rate (QBER) and (c) secure key rate (SKR) over time for Alice and Bob.} Each data point stands for a 10 s average value.
\end{figure*}

To verify the feasibility of the network, we measure the two-photon coincidences for two representative phase settings as shown in FIG.\ref{Fig3}, i.e., $ \phi_{s} + \phi_{i} = 0 $ in FIG.\ref{Fig3}(a) and $ \phi_{s} + \phi_{i} = \pi $ in FIG.\ref{Fig3}(b). Each coincidence histogram has three peaks, revealing three relative arrival times due to the intrinsic nature of the Franson interferometer \cite{Franson1989}. The middle peak corresponds to the indistinguishable event, in which both photons travel along the long or the short paths. For this reason, the coincidences of central peaks vary remarkably between close to zero count ($ \phi_{s} + \phi_{i} = \pi $) and close to four times the side peaks ($ \phi_{s} + \phi_{i} = 0 $), whereas the side peaks have heights that are independent of phase. The time difference between the two arms of the Franson interferometer ($ \Delta T = 2.5 ~{\rm ns} $) is larger than the coherence time of the generated photons ($ \tau_{c} \sim 250 ~{\rm ps} $), and much smaller than the coherence time of the CW pump laser ($ \tau_{p} \sim 2.7 ~{\rm \mu s} $ for a typical linewidth $ \sim 60 ~{\rm kHz} $), which guarantees high interference visibility and, thus, high purity of energy-time entanglement.

Further, we measure the two-photon interference fringes by post-selecting each peak with a coincidence window of $ 2.5 ~{\rm ns} $. The results are shown in FIG.\ref{Fig4}(a)-(f) as a function of the phase $ \phi_{s} + \phi_{i} $ for both central and side peaks in the coincidence histograms. One can clearly see that the central peak oscillates sinusoidally with the phase, and no interference can be observed for the side peaks. The fitting results show that all the raw visibilities are above $ 0.856 $, which are greater than the classical bound ($ \sim 0.707 $), required for the violation of the CHSH form of the Bell's inequality \cite{Clauser1969}.  The detected brightness (including all the insertion losses of all components), is defined as detected photon pairs rate per squared milliwatt per megahertz. In our work, the average detected brightness for all user pairs is $ 0.52 \times 10^{-2} ~{\rm s^{-1} mW^{-2} MHz^{-1}} $, with a minimum of $ 0.40 \times 10^{-2} ~{\rm s^{-1} mW^{-2} MHz^{-1}} $ and a maximum of $ 0.78 \times 10^{-2} ~{\rm s^{-1} mW^{-2} MHz^{-1}} $, and details can be found in Appendix B.

We use the BBM92 protocol \cite{Bennett1992} to establish the keys between Alice and Bob. In FIG.\ref{Fig1}(e), we show the measurement setup. Alice/Bob splits their photons with a 50:50 beam splitter, which performs the random choice of measurement basis between \textit{Z} ($ 0 $ / $ \pi $) and \textit{X} ($ \frac{\pi}{2} $ / $ \frac{3\pi}{2} $). An additional circulator is added at the input of each interferometer to obtain two-port data from each interferometer. All eight outputs are monitored with SNSPDs. For the long-term operation of the whole setup, we have made some experimental technical efforts as described in Appendix D. For example, we use feedback control to move the pump wavelength to follow the resonance drift of the MRR resonance due to the thermal instability of the laboratory, which makes the resonance stable for more than 20 hours. FIG.\ref{Fig5} shows the quantum key distribution data between Alice and Bob in $ 12 000~{\rm s} $, which are measured between CH33 and 37 directly from the outputs of AWG. In FIG.\ref{Fig5}(a), we show the entanglement visibility and obtain an average value of $ 93.84 \pm 0.61 \% $. The average quantum bit error rate (QBER) is $ 3.08 \pm 0.30\% $, as shown in FIG.\ref{Fig5}(b), which is calculated by $ QBER = (1-Visibility)/2 $ \cite{Gisin2002}. The secure key rate (SKR) is calculate using $ SKR \geq n_{\mathrm{sift}}\left[1-f\left(\delta_{b}\right) H_{2}\left(\delta_{b}\right)-H_{2}\left(\delta_{p}\right)\right] $ \cite{Ma2007, Yin2017}, where $ n_{\mathrm{sift}} $ is the sifted key rate, $ \delta_{b} (\delta_{p}) $ the bit (phase) error rate, $ f\left(x\right) $ is the error correction efficiency as a function of error rate (here we set $ f\left(x\right) = 1.2 $), and $ H_{2}\left(x\right) $ is the binary entropy function: $ H_{2}(x)=-x \log _{2} x-(1-x) \log _{2}(1-x) $. Due to the symmetry of X- and Z-basis measurements, $ \delta_{b} = \delta_{p} = E_{\lambda} $, where $ E_{\lambda} $ is the overall QBER. Within an integration time of $ 12 000~{\rm s} $, we obtain a total security key of $ 2.46 \times 10^{6} $ bits. This corresponds to an average SKR of $ 205 ~{\rm bits/s} $, as shown in FIG.\ref{Fig5}(c). We have measured the quantum key distribution in $ 2000~{\rm s} $ with all six pairs of entangled photons and present the data in Appendix C.

\section{CONCLUSION}
In conclusion, we generate comb-like bipartite energy-time entangled states in a chip-integrated Si$_{3}$N$_{4}$ (MRR), and experimentally demonstrated a fully connected four-user tabletop qn by multiplexing 12 DWDM channels such that correlation exists between every possible two-party link. The broadband and comb-like spectrum combined with considerable brightness, make the Si$_{3}$N$_{4}$ entangled source highly promising for large-scale, wavelength-multiplexed, fully connected QNs and flexible to interface to other integrated quantum photonics applications.

Based on our measured single-photon spectrum, a total of 44 wavelength channels can be used to build a fully connected quantum communication network with at least 7 users. To further expand the scale of network users requires more wavelength channels and the use of optical splitters or other optical components. In addition, with the recent advancements of SiN fabrication technology, it is possible to integrate wavelength demultiplexing/multiplexing module onto the same chip of the source, which will largely reduce the complexity and the optical connection losses of the off-chip, individual optical components. This work paves the way for providing a turn-key solution of the large-scale entanglement-based QN.

\begin{acknowledgements}
	This research is supported by the National Key Research and Development Program of China (2017YFA0303704 and 2019YFA0308700), National Natural Science Foundation of China (Grants No.11690032, No.11321063), NSFC-BRICS (No.61961146001), Leading-edge Technology Program of Jiangsu Natural Science Foundation (BK20192001), and the Fundamental Research Funds for the Central Universities.
\end{acknowledgements}

\appendix

\renewcommand\thefigure{\Alph{section}S\arabic{figure}}
\renewcommand\thetable{\Alph{section}S\arabic{table}}
\setcounter{figure}{0}

\section{CHARACTERIZATION OF MICRORING RESONATORS}
\begin{figure}[htbp]
	\centering
	\includegraphics[width=0.45\textwidth]{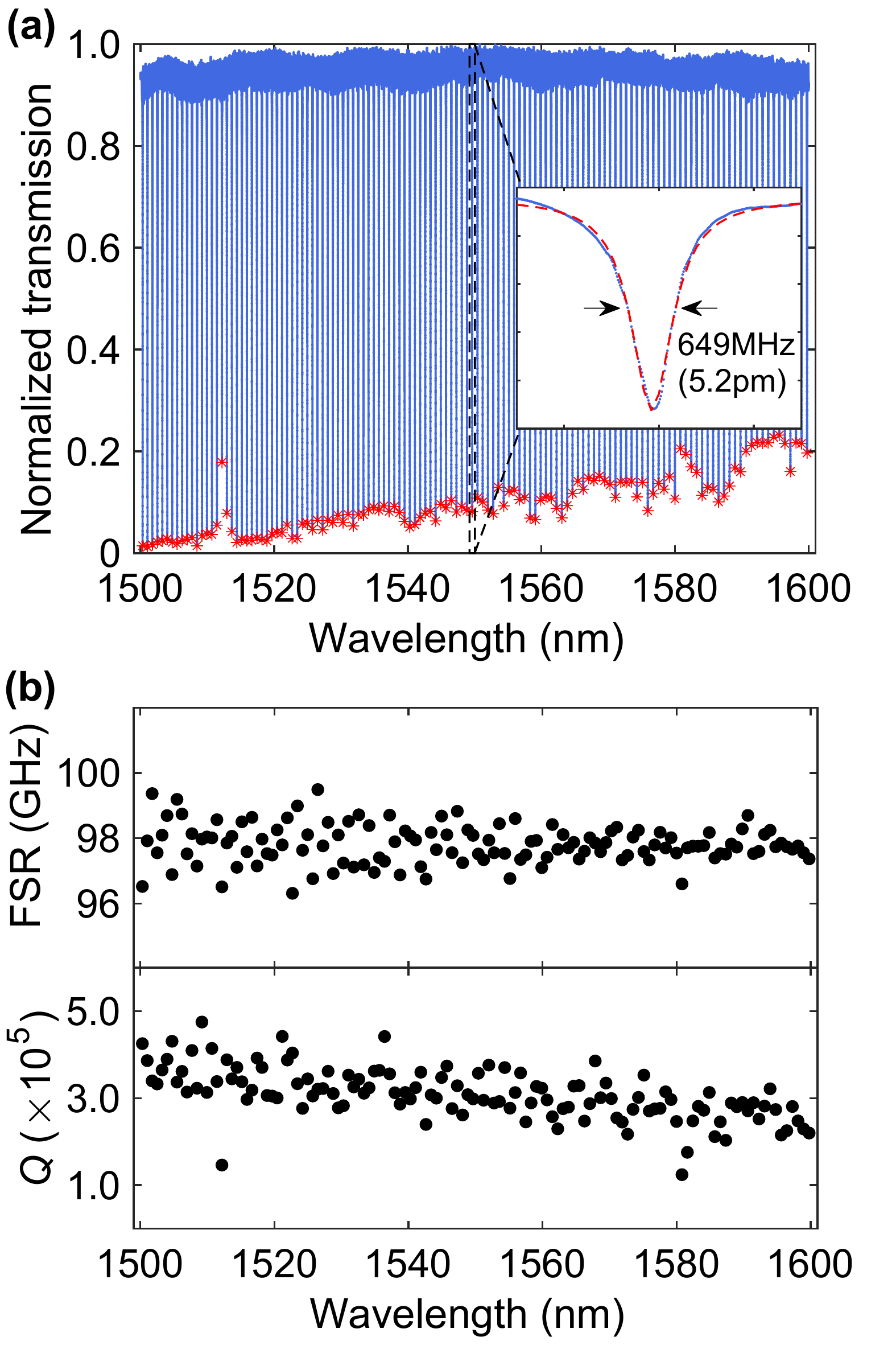}
	\caption{\label{FigS1}Characterization of the MRR with laser. (a) The measured cavity transmission spectrum ranging from $ 1500 $ to $ 1600 ~{\rm nm} $. The inset shows the resonance spectra near the pump. (b) The free spectral range (FSR) and estimated quality (\textit{Q}) factor of the resonator.}
\end{figure}

To characterize the MRR, we measure the transmission spectrum ranging from $ 1500 $ to $ 1600 ~{\rm nm} $, as shown in FIG.\ref{FigS1}(a).

The inset shows an enlarged view of a resonant mode around the pump wavelength (CH35) with a FWHM of $ 5.2 ~{\rm pm} $, or $ 649 ~{\rm MHz} $. FIG.\ref{FigS1}(b) shows the FSR and the quality factors (\textit{Q}) of each resonant mode. We obtain $ 128 $ cavity resonance modes and the averaged FSR is $  97.8 ~{\rm GHz} $, which is sufficiently close to the standard $  100 ~{\rm GHz} $ DWDM filters and offers a plentiful selection of possible photon pairs at telecom-band. The averaged measured \textit{Q} factor is about $ 3.1 \times 10^{5} $.

\begin{table*}[ht]
	\begin{tabular}{lcccccc}
		\hline
		\multirow{2}{*}{Channel piars} & Coherence time & Bandwidth & PGR  & PGR  & Loss   & Jitter \\
		& (ps)           & (MHz)     & (s$ ^{-1} $mW$ ^{-2} $)  & (s$ ^{-1} $mW$ ^{-2} $MHz$ ^{-1} $ )  & (dB)   & (ps)   \\ \hline
		CH37-CH33                      & 250.8          & 634.7     & $ 7.22\times10^{3} $ & 11.4 & -10.29 & 138.3  \\
		CH38-CH32                      & 234.4          & 678.9     & $ 8.70\times10^{3} $ & 12.8 & -11.22 & 133.9  \\
		CH39-CH31                      & 245.0          & 649.7     & $ 7.27\times10^{3} $ & 11.2 & -10.82 & 136.4  \\
		CH40-CH30                      & 254.6          & 625.2     & $ 9.55\times10^{3} $ & 15.3 & -11.00 & 127.0  \\
		CH41-CH29                      & 242.2          & 657.1     & $ 7.92\times10^{3} $ & 12.1 & -11.12 & 139.8  \\
		CH42-CH28                      & 244.6          & 650.7     & $ 9.51\times10^{3} $ & 14.6 & -11.85 & 128.1  \\ \hline
	\end{tabular}
	\caption{\label{TabS1}The coherence time of the single photons, bandwidth, pair generation rate (PGR), calculated loss and calculated time jitter of all six selected photon pairs without interferometers and partial WDM network.}
\end{table*}

\begin{figure}[htbp]
	\centering
	\includegraphics[width=0.45\textwidth]{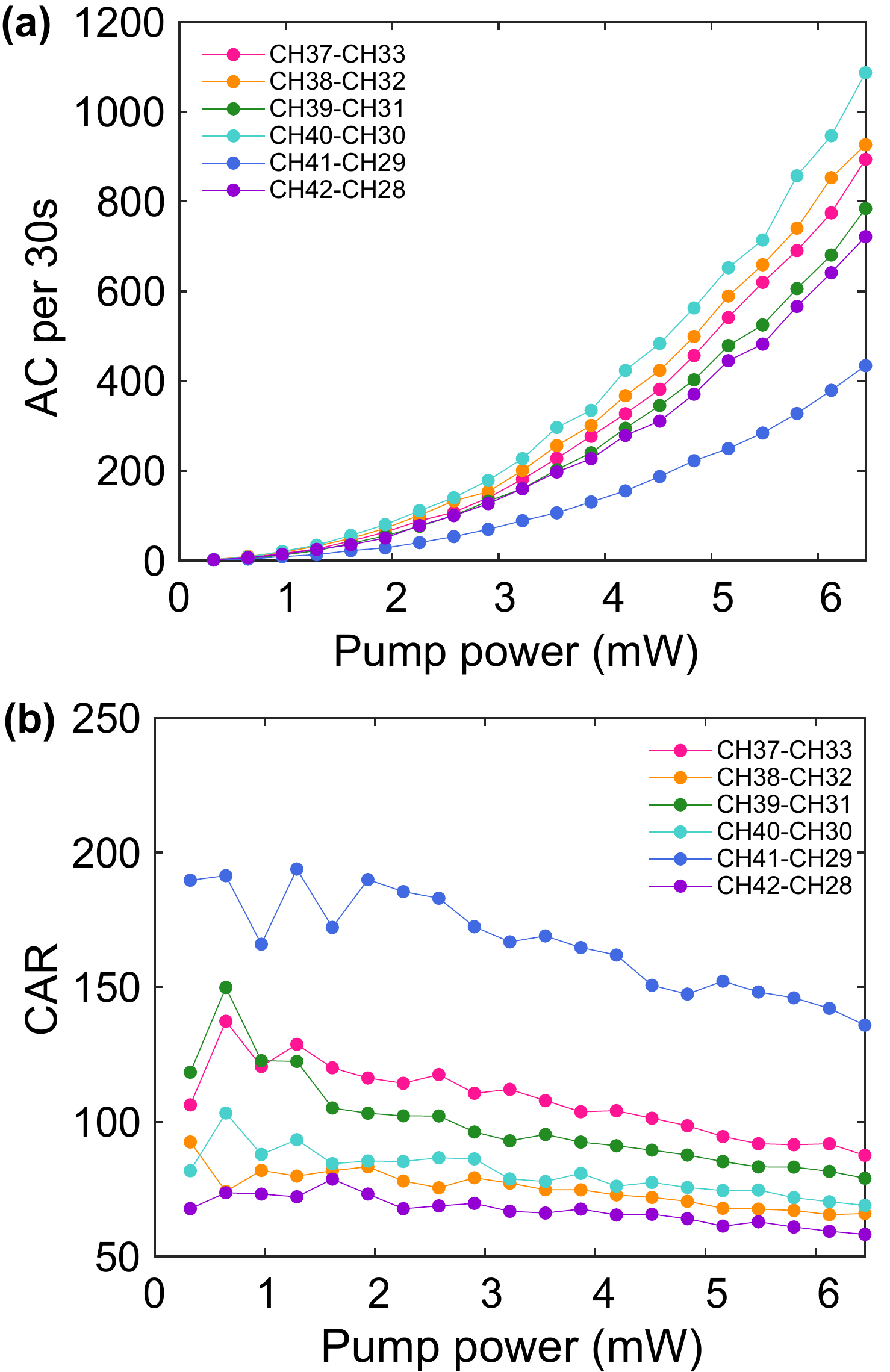}
	\caption{\label{FigS2}(a)Accidental coincidences (AC)  and (b) coincidence to accident ratio (CAR) as a function of the injected pump power for all six selected photon pairs.}
\end{figure}

\begin{figure}[htbp]
	\centering
	\includegraphics[width=0.45\textwidth]{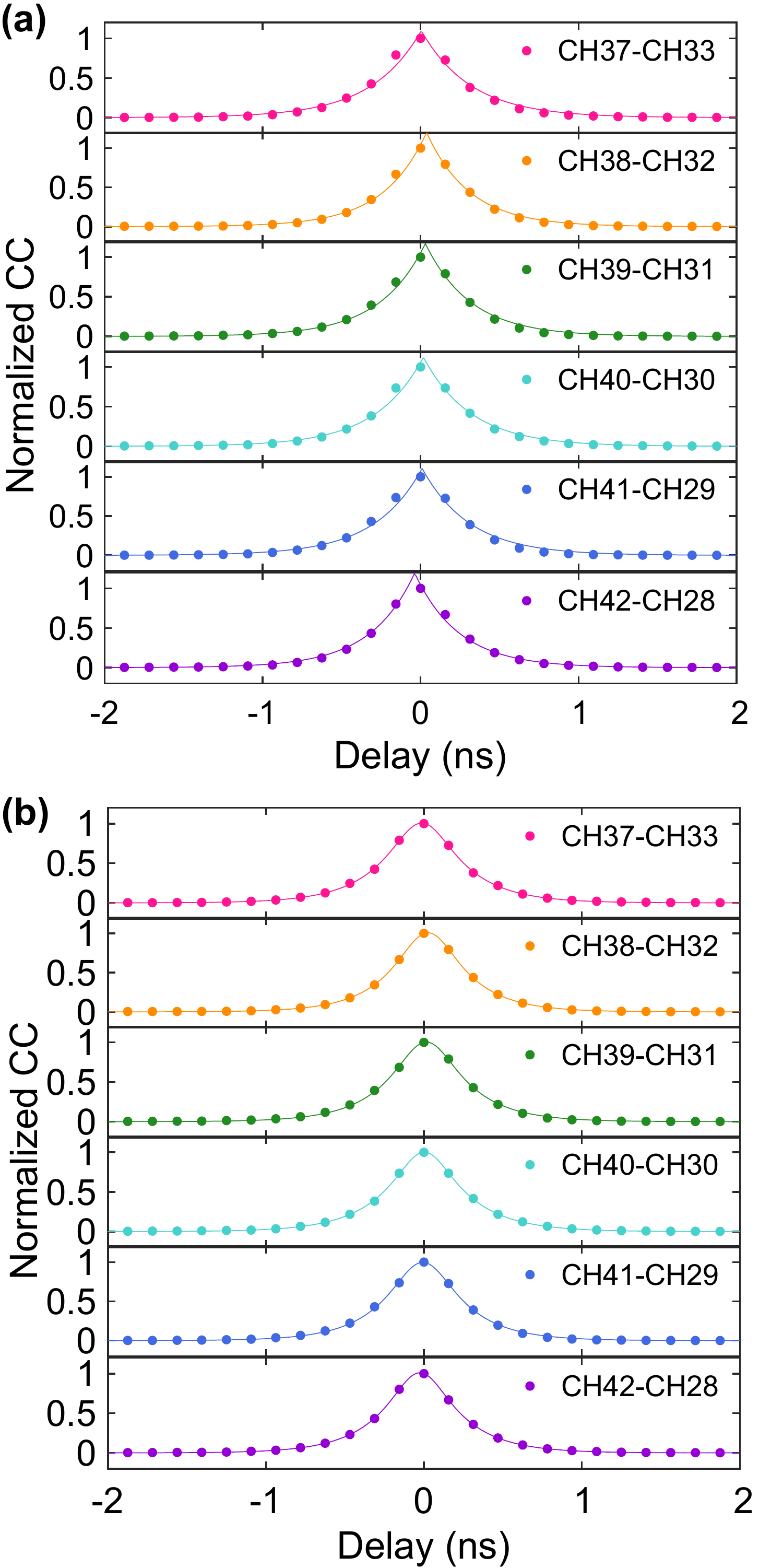}
	\caption{\label{FigS3}Comparison of the fitting results of using (a) $ \frac{A}{\tau_{c}} \exp \left( -\frac{\left| \Delta t \right|}{\tau_{c}} \right) $ and (b) $ \frac{A}{\tau_{c}} \exp \left( -\frac{\left| \Delta t \right|}{\tau_{c}} \right) \ast  \exp \left( -\frac{(\Delta t)^{2} }{\sigma^{2}} \right) $.}
\end{figure}

Then we characterized the photon pair generated from the MRR. The coincidence counts (CC) in FIG.\ref{Fig2}(b) is obtained by taking a window with a center at zero relative delay and a width of 2.5 ns from the histogram of coincidence counts (such as the histogram shown in the inset of FIG.\ref{Fig2}(b) with a temporal resolution of $ 156.25 ~{\rm ps} $), and summing all the counts in this window. Accidental coincidences (AC) is the average of the counts in all other 2.5ns windows except the previous one in the histogram. Coincidence-to-accidental ratio (CAR) is defined as CC divided by AC. FIG.\ref{FigS2} shows the AC and CAR as a function of injected pump power. At low pump power, the CAR is mainly limited by the dark count of the detector. As the pump power increases, the multipair generation increases, which leads to a reduction in the CAR. Taking the pair of channels CH37 and CH33 as an example, we obtain a maximum CAR of 137 when the pump power is $ 0.64~\rm{mW} $, and the CAR is 95 when the pump power is $ 5~\rm{mW} $.

The measured coincidence count curve is the convolution of the signal/idler temporal correlation function $ g_{s,i}^{\left( 2 \right)}\left( t \right) $  and the response function of the whole detection system $ h\left( t \right) $ \cite{Zhou2014}. The signal/idler temporal cross-correlation function is given as $ g_{s,i}^{\left( 2 \right)}\left( \Delta t \right) \propto \frac{1}{\tau_{c}} \exp \left( -\frac{\left| \Delta t \right|}{\tau_{c}} \right) $ \cite{Reimer2014}, and the response of the whole detection system can be modeled using a Gaussian function $ h\left( \Delta t \right) \propto \exp \left( -\frac{(\Delta t)^2}{\sigma ^2} \right) $ \cite{Zhou2014}. Here $ \Delta t $ is the time delay, $ \tau_{c} $ is the coherence time of the single photons and $ \sigma $ is the time jitter of the detection system (including SNSPD and time-tagger unit). In FIG.\ref{FigS3}, we compare the fitting results without (a) and with (b) considering the broadening due to the jitter of  the detection system. After considering the time jitter, we obtained an average coherence time of $ 245.2 ~\rm{ps} $ and an average bandwidth of $ 649.4 ~\rm{MHz} $, which is in good agreement with the linewidth of the MRR. Detailed data of all six pairs of photons are listed in TABLE.\ref{TabS1}.

The PGR can be calculated by $ PGR=\frac{S_sS_i}{R_c} $, and the loss can be derived by $ Loss=\frac{R_c}{\sqrt{S_sS_i}} $, where $ S_{s} $ ($ S_{i} $) is the single counts of signal (idler) due to SFWM, and $ R_{c} $ is the coincidence count. It can be found that the calculated PGR is different for different channel pairs, which may come from the nonideal dispersion engineering and high-order mode-crossing \cite{Ji2021} of MRR. This can also be seen in the transmission spectrum of MRR with nonuniform extinction ratio in FIG.\ref{FigS1}(a). We estimate an average PGR for each pair of photons to be $ 8.3\times10^{3} ~s^{-1}mW^{-2} $ (ranging from $ 7.22\times10^{3} ~s^{-1}mW^{-2} $ to $ 9.55\times10^{3} ~s^{-1}mW^{-2} $ for different pairs), or $ 12.9 ~s^{-1}mW^{-2}MHz^{-1} $ (ranging from $ 11.2 ~s^{-1}mW^{-2}MHz^{-1} $ to $ 15.3 ~s^{-1}mW^{-2}MHz^{-1} $ for different pairs) after dividing by the bandwidth, as listed in the TABLE.\ref{TabS1}. The PGR of different channel pairs together with the different insertion loss of different wavelength channels, resulting in the different coincidence counts of the six channel pairs show in FIG.2(b).

\begin{table*}[ht]
	\begin{tabular}{lcccccc}
		\hline
		\multicolumn{1}{c}{\multirow{2}{*}{User}} & \multirow{2}{*}{ITU   Channels} & Detected   Brightness       & \multicolumn{2}{c}{Total Loss (dB)} & \multicolumn{2}{c}{Visibility} \\
		\multicolumn{1}{c}{}                      &                                 & (s$ ^{-1} $mW$ ^{-2} $MHz$ ^{-1} $) & Signal            & Idler             & Raw            & Net           \\ \hline
		Alice \& Bob                              & CH37-CH33                     & $ 0.78\times10^{-2} $       & -14.29            & -13.20            & $ 92.70\pm0.70\% $  & $ 99.53\pm0.17\% $ \\
		Alice \& Chloe                            & CH38-CH32                     & $ 0.58\times10^{-2} $       & -14.90            & -13.12            & $ 90.50\pm0.90\% $  & $ 97.67\pm0.44\% $ \\
		Alice \& Dave                             & CH39-CH31                     & $ 0.40\times10^{-2} $       & -15.27            & -15.30            & $ 85.56\pm1.37\% $  & $ 96.70\pm0.65\% $ \\
		Bob   \& Chloe                            & CH40-CH30                     & $ 0.46\times10^{-2} $       & -14.03            & -14.01            & $ 91.89\pm0.94\% $  & $ 99.83\pm0.13\% $ \\
		Bob   \& Dave                             & CH41-CH29                     & $ 0.47\times10^{-2} $       & -13.86            & -14.67            & $ 91.25\pm0.97\% $  & $ 98.93\pm0.32\% $ \\
		Chloe \& Dave                             & CH42-CH27                     & $ 0.44\times10^{-2} $       & -14.29            & -14.56            & $ 89.48\pm1.10\% $  & $ 96.87\pm0.59\% $ \\ \hline
	\end{tabular}
	\caption{\label{TabS2}The brightness detected between users, the total loss of different wavelength channels, and the corresponding raw (net) visibilities without (with) subtraction of accidental coincidences.}
\end{table*}

\section{RESULTS OF ENTANGLEMENT DISTRIBUTION FOR SIX PAIRS OF USERS}

In order to conveniently determine the phase $ \phi_{s} + \phi_{i} $, the two-photon interference fringes shown in Fig.4(a)-(f) is measured by keep the idler's phase zero while tuning the phase of the signal. The results show that the visibility (calculated from raw data) of different user pairs varies from $ 85.56\pm1.37\% $ (Alice-Dave) to $ 92.70\pm0.70\% $ (Alice-Bob). We tested the total loss from chip to SNSPD for each wavelength channel with classical laser and power meter, and the results showed that the higher the total loss of signal and idler, the lower the interference visibility between the corresponding user pairs. Note that each user received three wavelength channels and shared one and only one pair of correlated photons with any other user, and coincidences between uncorrelated photons contributes to the accidental coincidences. After subtracting the measured accidental coincidence, the net visibilities are found to be varies from $ 96.70\pm0.65\% $ to $ 99.83\pm0.13\% $, as listed in TABLE.\ref{TabS2}.

\begin{figure}[h]
	\centering
	\includegraphics[width=0.45\textwidth]{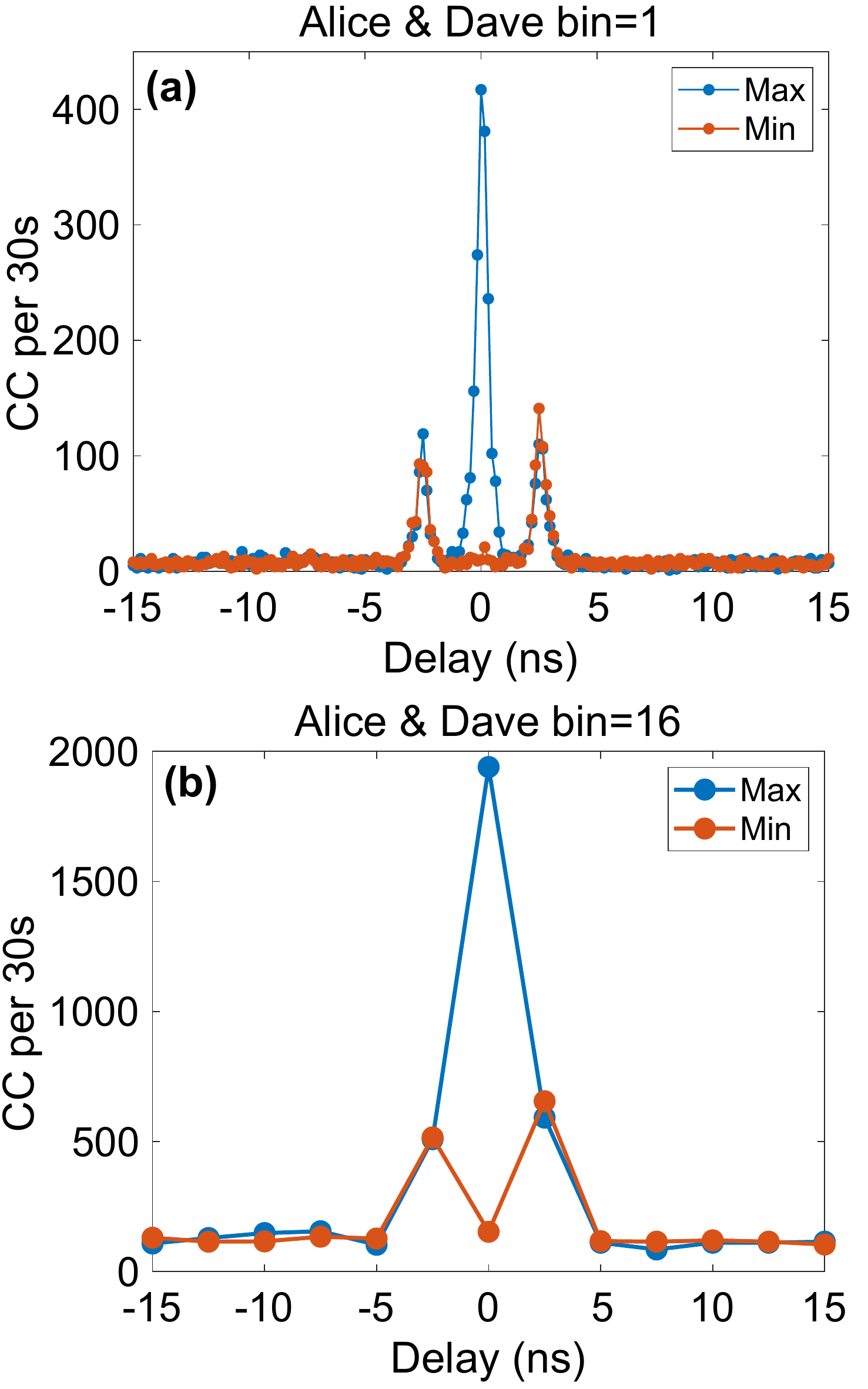}
	\caption{\label{FigS4}The coincidence histograms of the maximum (blue) and minimum (orange) points of the interference curve between Alice and Dave, with the time-bin size of (a) is 0.15625ns, and the time-bin size of (b) is 2.5ns.}
\end{figure}

Let us take the interference curve between Alice and Dave shown in FIG.\ref{Fig1}(c) as an example, and the corresponding coincidence histograms of the maximum (blue) and minimum (orange) points are shown in FIG.\ref{FigS4}. The left and right figures are the same set of data, with the time-bin size of (a) being $ 0.15625 ~\rm{ns} $ and the time-bin size of (b) being $ 2.5 ~\rm{ns} $. The middle point of the two curves in (b) shows a maximum count of 1940 and a minimum count of 151, which corresponds to a raw visibility of $ 85.6\% $. The side peaks correspond to the non-interfering terms in typical Franson interferometer (see main text for details). The accidental coincidence is obtained by taking the average of the other 10 points except the 3 middle points. The averaged accidental coincidence counts calculated by the two curves are 117 and 120, respectively. After subtracting the respective accidental coincidences, we obtain a net visibility of $ 96.7\% $, proving the presence of high-quality Bell nonlocality between users. However, we would like to emphasize that in the later BBM92 QKD experiment, no accidental coincidence counts were subtracted.

\begin{figure}[h]
	\centering
	\includegraphics[width=0.45\textwidth]{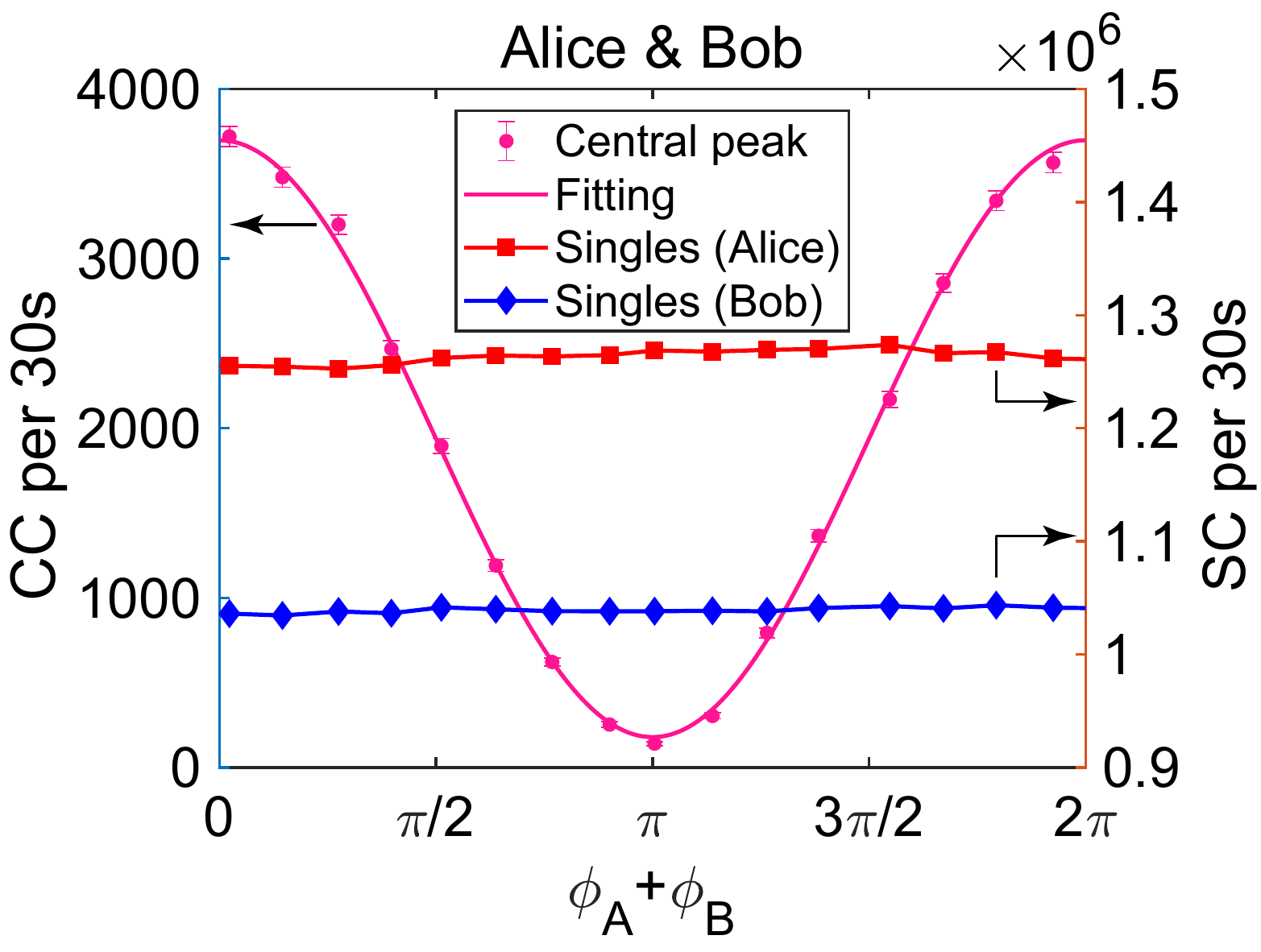}
	\caption{\label{FigS5}Coincidence count of the central peak and single count of the output of each interferometer as a function of the phase $ \phi_{s} + \phi_{i} $ for Alice and Bob.}
\end{figure}

FIG.\ref{FigS5} shows the coincidence count and single count of Alice and Bob as a function of the phase $ \phi_{A} + \phi_{B} $, where the fluctuation of single counts is less than $ 0.8\% $. The comparison demonstrates that single-photon coherence is suppressed because the time difference between the two arms of the Franson interferometer ($ \Delta T = 2.5 ~{\rm ns} $) is larger than the coherence time of the single photons ($ \tau_{c} \sim 250 ~{\rm ps} $).

\section{RESULTS OF QUANTUM KEY DISTRIBUTION WITHOUT PID MODULE}

\begin{figure}[h]
	\centering
	\includegraphics[width=0.45\textwidth]{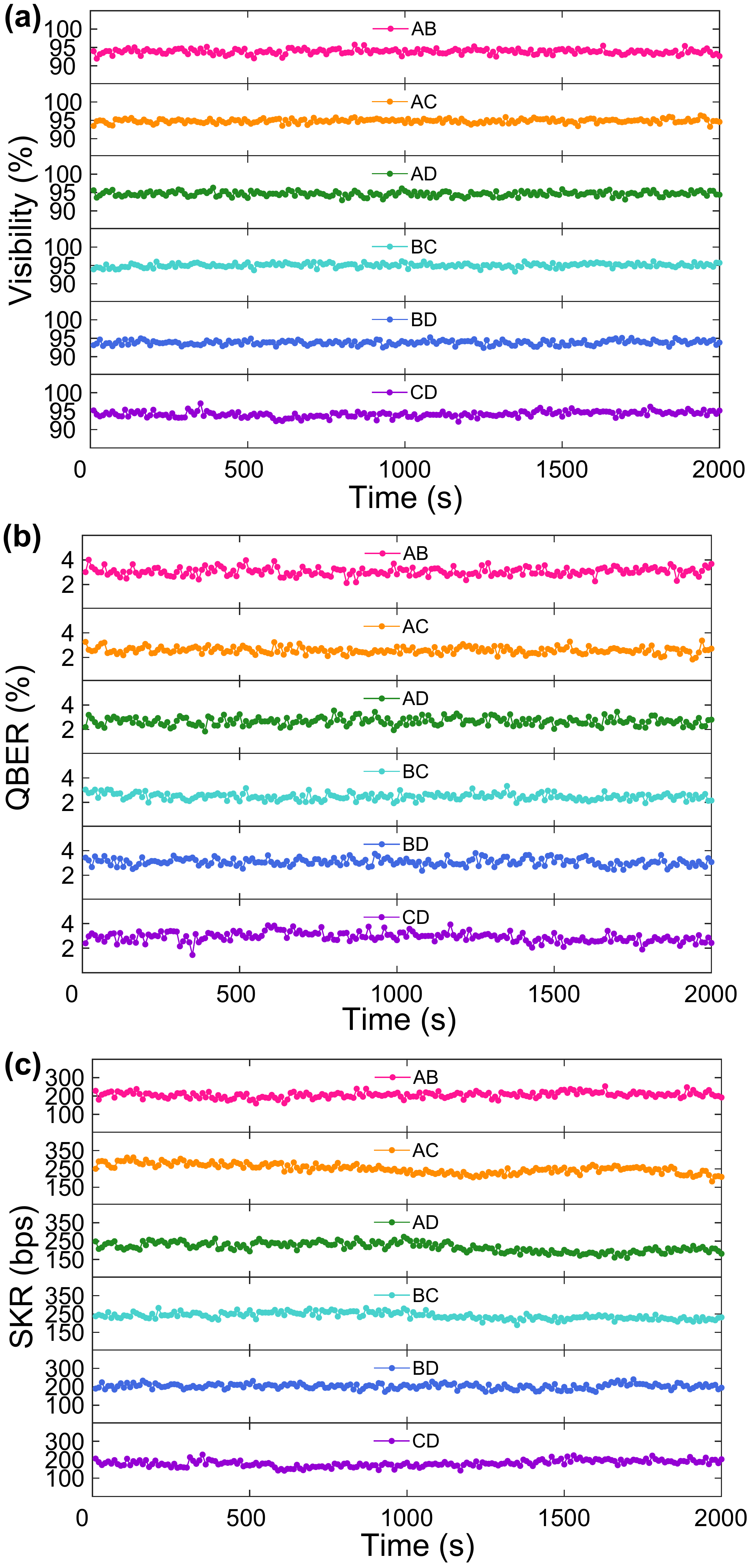}
	\caption{\label{FigS6}The stability of (a) visibility, (b) quantum bit error rate (QBER), and (c) SKR over time without the use of PID module.}
\end{figure}

\begin{table*}[ht]
	\begin{tabular}{lcccc}
		\hline
		\multicolumn{1}{c}{\multirow{2}{*}{User}} & Total sifted key & \multirow{2}{*}{Visibility} & \multirow{2}{*}{QBER} & SKR \\
		\multicolumn{1}{c}{}                      & ($ 2000~s$)          &                             &                       & (bit/s)         \\ \hline
		Alice \& Bob                              & $ 7.28\times10^{5} $             & $ 93.89\pm0.66\% $                     & $ 3.06\pm0.33\% $                & $ 206 $             \\
		Alice \& Chloe                            & $ 8.12\times10^{5} $             & $ 94.85\pm0.53\% $                    & $ 2.57\pm0.26\% $                & $ 252 $             \\
		Alice \& Dave                             & $ 7.13\times10^{5} $             & $ 94.65\pm0.64\% $                     & $ 2.67\pm0.32\% $                & $ 217 $             \\
		Bob \& Chloe                              & $ 7.60\times10^{5} $              & $ 95.04\pm0.53\% $                    & $ 2.48\pm0.27\% $                & $ 240 $             \\
		Bob \& Dave                               & $ 7.22\times10^{5} $             & $ 93.82\pm0.59\% $                     & $ 3.09\pm0.29\% $                & $ 203 $             \\
		Chloe \& Dave                              & $ 6.18\times10^{5} $             & $ 94.18\pm0.77\% $                     & $ 2.91\pm0.38\% $                & $ 180 $            \\ \hline
	\end{tabular}
	\caption{\label{TabS3}The measured total sifted key, the average interference visibility during key distribution, the average quantum bit error rate (QBER), and the average secure key rate (SKR) between four users in $ 2000~{\rm s} $.}
\end{table*}

We tested 2000 seconds of quantum key distribution (QKD) between all six pairs of entangled photon (or channels). FIG.\ref{FigS6} shows the overall qubit error rate (QBER) between $ 2.57 \% $ and $ 3.09\% $, and the averaged SKR between $ 180 $ and $ 251 ~{\rm Hz} $. The results are summarized in TABLE.\ref{TabS3}. The imperfection of the visibility mainly comes from the limited extinction ratio and isolation of the WDM filter, the phase drifts of the interferometer, and the imperfect polarization adjustment.

\section{{STABILITY OF OUR SYSTEM}}

\begin{figure}[hbp]
	\centering
	\includegraphics[width=0.3\textwidth]{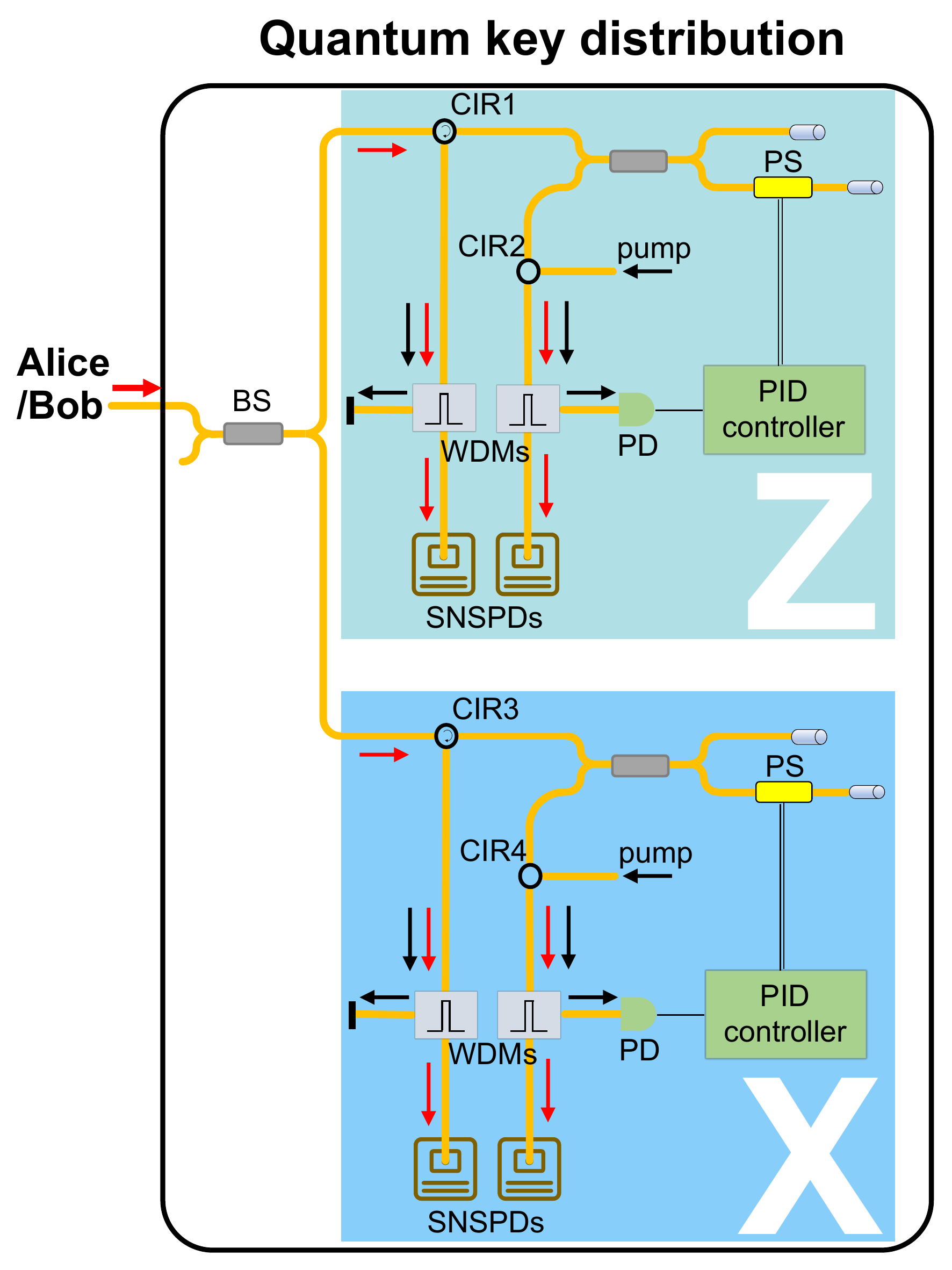}
	\caption{\label{FigS7}The measurement setup for quantum key distribution using BBM92 protocol with PID module included. \textit{Z} and \textit{X} represent two sets of measurement bases \textit{Z} ($ 0 $ / $ \pi $) and \textit{X} ($ \frac{\pi}{2} $ / $ \frac{3\pi}{2} $). The black arrows represent the direction of the reference light, and the red arrows represent the direction of the signal/idler. Abbreviations of components: BS, beam splitter; PD, power detector; PS, phase shifter; CIR, circulator; WDM, wavelength division multiplexer; SNSPD, superconducting nanowire single-photon detector.}
\end{figure}

\begin{figure}[h]
	\centering
	\includegraphics[width=0.45\textwidth]{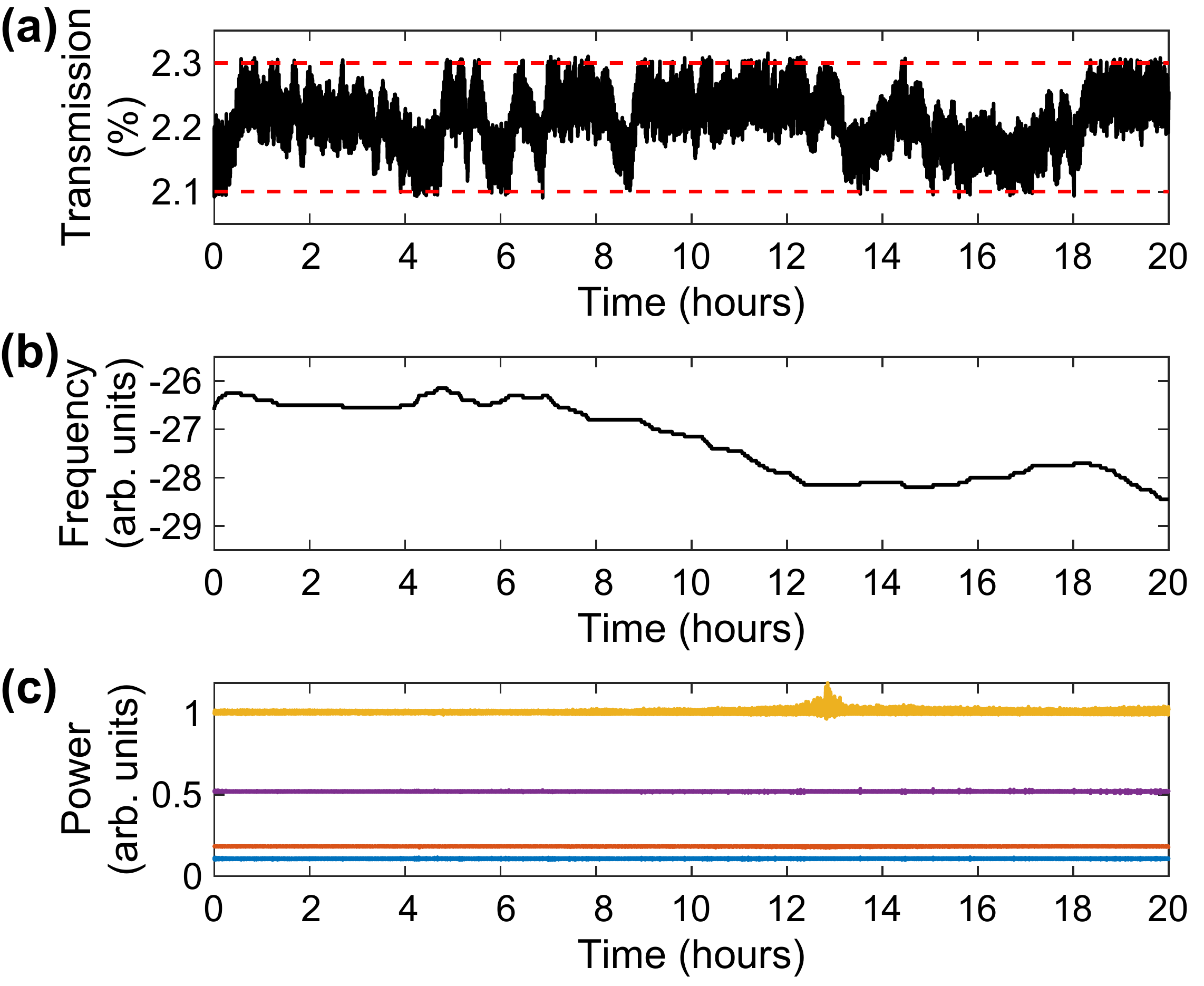}
	\caption{\label{FigS8}(a) The stability of pump transmission at resonance wavelength, (b) the tuning of pump frequency during feedback control, and (c) the stability of the output pump power (phase) of the four PID phase-locking modules used in quantum key distribution.}
\end{figure}

\begin{figure}[htbp]
	\centering
	\includegraphics[width=0.45\textwidth]{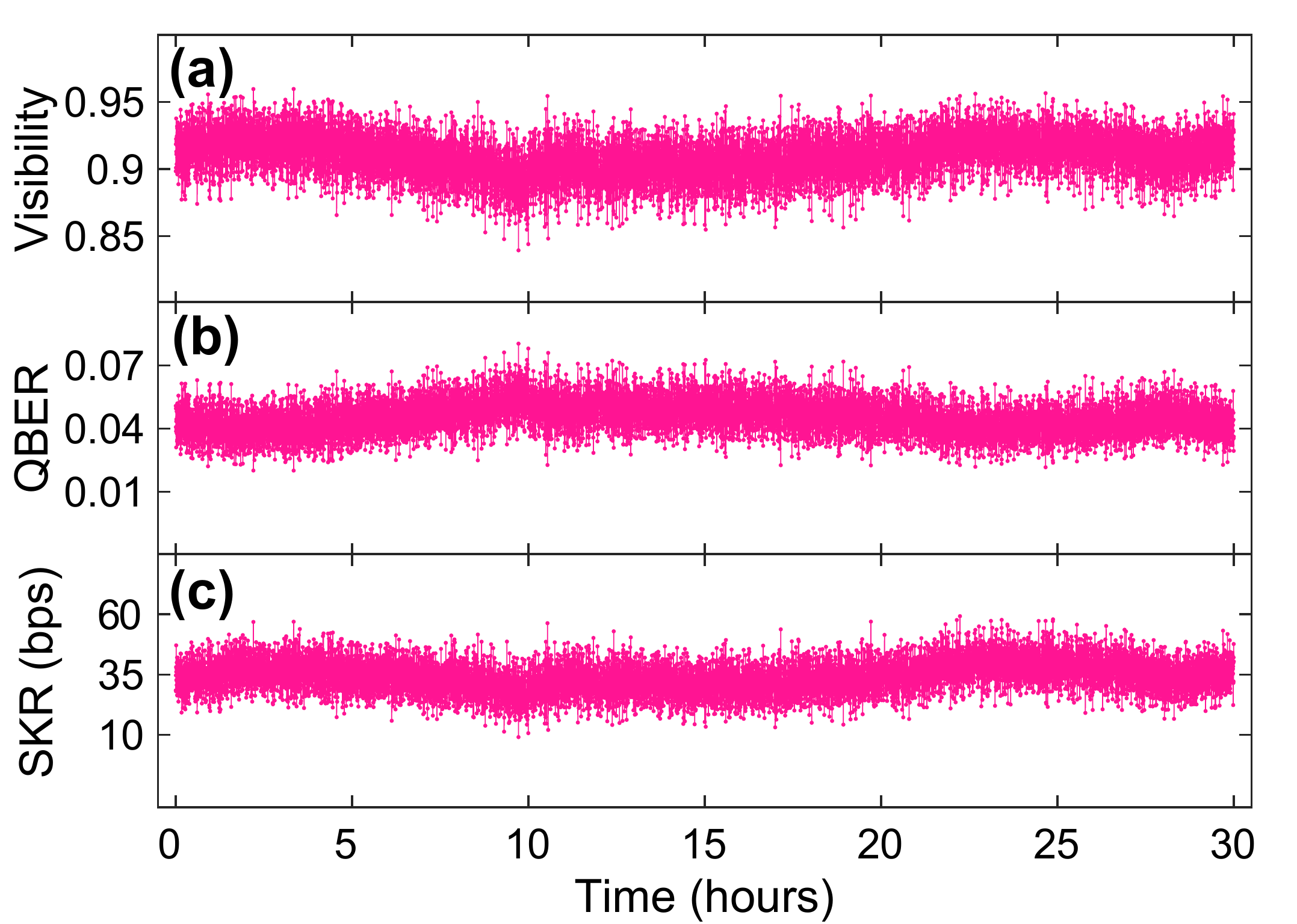}
	\caption{\label{FigS9}(a) Visibility, (b) quantum bit error rate (QBER) and (c) secure key rate (SKR) over time for Alice and Bob with PID phase-locking module included. Each data point stands for a 10 s average value.}
\end{figure}

Although all interferometers are entirely buried in quartz sand and wrapped with a thick insulating layer, the phase of the interferometer will still drift slightly due to the sensitivity of the phase to temperature, affecting the duration of QKD. For the long-term operation of the whole setup,  we have made some experimental technical efforts. We added PID phase-locking module based on the original interferometers as shown in FIG.\ref{FigS7}. Specifically, a portion of the pump from FIG.\ref{Fig1}(a) is split by beam splitter and then injected into the interferometers through a circulator as reference light for actively stabilizing the phase by using a PID controller. An additional circulator is added to the input of each interferometer to acquire two-port data. The excess pump photons are filtered by the WDMs and the desired signal/idler photons are selected for photon detection. The reflected pump power is monitored by a power detector whose output electrical signal is connected to a PID controller (Toptica DigiLock-110/PID-110), which is used to adjust the phase shifter, and actively stabilizes the phase of the interferometer. Alice and Bob have a total of four interferometers, so a total of four such PID modules are required. We define the initial phase of Alice's first interferometer as the zero phase point, whereas the phases of the other three interferometers are determined by their respective interference results. After that, the relative phase of the measurement base \textit{Z} ($ 0 $ / $ \pi $) and \textit{X} ($ \frac{\pi}{2} $ / $ \frac{3\pi}{2} $) are determined and then locked by the PID modules.

At the same time, to maintain the stability of the MRR resonance, we added a thermoelectric cooler and NTC thermistor at the bottom of the chip, and both were connected to a temperature controller (Thorlabs TED4015). However, due to the instability of the temperature of the laboratory, we found that the resonance of the MRR will shift slightly over time. To solve this problem, we developed a feedback control system to move the pump wavelength to follow the resonance drift of the MRR resonance. In the upper panel of FIG.\ref{FigS8}, we show the pump transmission at the resonant wavelength, which is kept at the minimum level (~$ 2.2\pm0.1\% $ of the input power, and the upper and lower bounds are marked with red dashed lines). In order to maintain this resonance condition, we adaptively tune the pump frequency, as shown in the middle panel of FIG.\ref{FigS8}. The range of the y-axis is about $ 150~\rm{MHz} $. The lower panel of FIG.\ref{FigS8} shows the variation of the pump power output from the four interferometers with time under the active control of the PID phase-locking module, which indicates that the phases of the four interferometers have good stability.

FIG.\ref{FigS9} shows the long-running performance of quantum key distribution between Alice and Bob. In FIG.\ref{FigS9}(a) we show the entanglement visibility with an average value of $ 90.98 \pm 1.61 \% $, and in FIG.\ref{FigS9}(b) the QBER with an average value $ 4.51 \pm 0.80\% $. Within an integration time of $ 30~{\rm h} $, we obtain a total security key of $ 3.68 \times 10^{6} $ bits. This corresponds to an average SKR \cite{Ma2007, Yin2017} of $ 34 ~{\rm bits/s} $, as shown in FIG.\ref{FigS9}(c). Note that due to the residual pump photons, the QBER is higher than that when the PID module is not added. At the same time, due to the extra loss brought by WDMs, the SKR is lower than that without the PID module included.

\section{{SECURITY OF OUR SYSTEM}}
Entanglement-based QKD has inherent source-independent security and is therefore immune to source-targeted attacks \cite{Ma2007, Koashi2003}. Therefore, one usually considers the effect of detection attacks for entanglement-based QKD, such as detector-related attack \cite{Zhao2008, Lydersen2010, Weier2011}, wavelength-dependent attack \cite{Li2011}, spatial-mode attack \cite{Sajeed2015}. Note that most of these attacks have been studied in Ref.\cite{Yin2020}. Here we only consider the potential attacks coming from different physical implementations between ours and those in Ref.\cite{Yin2020}. For the detector efficiency mismatch attack, we use SNSPDs in our system and they are in work the free-running mode. Hence Eve cannot employ the detector efficiency mismatch attack. For the blinding attack, it has been shown that by implementing an active antilatching system for SNSPDs, monitoring the comparator pulse shape and implementing postprocessing key extraction, one can resolve the blinding attack \cite{Elezov2019}. For the beam splitter attack, we do use a 1×2 beam-splitter to randomly select the measurement basis in our QKD system. However, the photons generated by our MRR source inherently have narrow linewidths ($ \sim 650{\rm MHz} $), so the wavelength dependence of the beam splitter’s splitting ratio within this range can be ignored.

\section{{COMPARISON WITH OTHER QUANTUM NETWORKS BASED ON BBM92 PROTOCOL}}

\begin{table*}[htbp]
	\begin{tabular}{ccccccc}
		\hline
		\multirow{2}{*}{ }                                         & \multirow{2}{*}{Platform} & \multirow{2}{*}{Material} & \multirow{2}{*}{Source size}                              & \multirow{2}{*}{\begin{tabular}[c]{@{}c@{}}CMOS\\ compatible?\end{tabular}} & \multirow{2}{*}{\begin{tabular}[c]{@{}c@{}}Linewidth per \\frequency mode (GHz) \end{tabular}} & \multirow{2}{*}{\begin{tabular}[c]{@{}c@{}}Spectrum\\ bandwidth (nm) \end{tabular}} \\
		&                           &                           &                                                           &                                                                             &                                                                                               &                                                                                        \\ \hline
		\begin{tabular}[c]{@{}c@{}}Wengerowsky et al.\\ \cite{Wengerowsky2018}\end{tabular} & Free-space                & MgO:PPLN                  & \begin{tabular}[c]{@{}c@{}}4 cm\\ long\end{tabular}       & No                                                                          & 100                                                                                           & \begin{tabular}[c]{@{}c@{}}$\sim$60\\ (Fig.3 of ref.\cite{Wengerowsky2018})\end{tabular}                   \\
		\begin{tabular}[c]{@{}c@{}}Joshi et al.\\ \cite{Joshi2020}\end{tabular}       & Free-space                & MgO:PPLN                  & \begin{tabular}[c]{@{}c@{}}4 cm\\ long\end{tabular}       & No                                                                          & 100                                                                                           & \begin{tabular}[c]{@{}c@{}}$\sim$60\\ (SM of ref.\cite{Joshi2020})\end{tabular}                      \\
		\begin{tabular}[c]{@{}c@{}}Fitzke et al.\\ \cite{Erik2022}\end{tabular}      & Fiber-coupled             & PPLN                      & N.A.                                                      & No                                                                          & \begin{tabular}[c]{@{}c@{}}25, 50, 100\end{tabular}                              & \begin{tabular}[c]{@{}c@{}}$\sim$75\\ (Section A of ref.\cite{Erik2022})\end{tabular}               \\
		This work                                                       & Chip-integrated           & Si$_3$N$_4$                       & \begin{tabular}[c]{@{}c@{}}460 um\\ diameter\end{tabular} & Yes                                                                         & 0.65                                                                                          & \begin{tabular}[c]{@{}c@{}}$\sim$35\\ (Hundreds in ref.\cite{Reimer2016, Marin2017, Samara2019})\end{tabular}      \\ \hline
	\end{tabular}
	\caption{\label{TabS4}Comparison of sources used in different multi-user QKD platforms based on BBM92 protocol.}
\end{table*}

\begin{table*}[htbp]
	\begin{tabular}{cccccc}
		\hline
		\multirow{2}{*}{ }                                     & \multirow{2}{*}{Protocol} & \multirow{2}{*}{Basis} & \multirow{2}{*}{\# of users}                                     & \multirow{2}{*}{SKR (bps)}                                                                                                             & \multirow{2}{*}{QBER}                                                                            \\
		&                           &                        &                                                                  &                                                                                                                                        &                                                                                                  \\ \hline
		\begin{tabular}[c]{@{}c@{}}Wengerowsky et al.\\ \cite{Wengerowsky2018}\end{tabular} & BBM92                     & polarization           & \begin{tabular}[c]{@{}c@{}}4\\ (fully connected)\end{tabular}    & \begin{tabular}[c]{@{}c@{}}3$\sim$15\\ (estimated by authors)\end{tabular}                                                         & N.A.                                                                                             \\
		\begin{tabular}[c]{@{}c@{}}Joshi et al.\\ \cite{Joshi2020}\end{tabular}       & BBM92                     & polarization           & \begin{tabular}[c]{@{}c@{}}8\\ (fully connected)\end{tabular}    & \begin{tabular}[c]{@{}c@{}}58$\sim$304 (in-laboratory)\\ 0.5$\sim$51.8 (city-wide)\\ (calculated from SM)\end{tabular} & \begin{tabular}[c]{@{}c@{}}3\%$\sim$6\%\\ (estimated from SM)\end{tabular}                       \\
		\begin{tabular}[c]{@{}c@{}}Fitzke et al.\\ \cite{Erik2022}\end{tabular}      & BBM92                     & Time-bin               & \begin{tabular}[c]{@{}c@{}}4\\ (pairwise connected)\end{tabular} & \begin{tabular}[c]{@{}c@{}}42/29 (in-laboratory)\\ 6/102 (field test)\end{tabular}                        & \begin{tabular}[c]{@{}c@{}}2.41\%/2.36\% (in-laboratory)\\ 4.5\%/2.6\% (field test)\end{tabular} \\
		This work                                                       & BBM92                     & Energy-time            & \begin{tabular}[c]{@{}c@{}}4\\ (fully connected)\end{tabular}    & 180$\sim$252 (in-laboratory)                                                                                                           & 2.5\%$\sim$3.1\%                                                                                 \\ \hline
	\end{tabular}
	\caption{\label{TabS5}Performance comparison of QKD systems on different multi-user QKD platforms based on BBM92 protocol.}
\end{table*}

Recently, several platforms that use wavelength division multiplexing schemes to realize multiuser QNs have been presented \cite{Wengerowsky2018, Joshi2020, Liu2020, Liu2022, Erik2022} and some of these platforms have already been implemented in real-world scenarios. Here, we compare the performance of sources used by different platforms based on the BBM92 protocol in TABLE.\ref{TabS4}, and compare the performance of the QKD system of these platforms in TABLE.\ref{TabS5}.

\bibliography{SiN-WDM-Bib}

\end{document}